\documentclass[footinbib,onecolumn,showpacs,amsmath,amstex,amssymb,mathfonts,superscriptaddress]{revtex4}
\usepackage{amsfonts}

\usepackage{graphicx,amsmath,amssymb}
\usepackage{epstopdf}
\usepackage{grffile}
\usepackage[usenames]{color}
\usepackage{indentfirst}
\usepackage{float}
\usepackage{color}
\usepackage{mathrsfs}
\usepackage{dcolumn}
\usepackage{bm}
\usepackage{extramarks,fancyhdr}

\begin{document}

\title{Possibility of superradiant phase  transitions in coupled two-level atoms }
\author{Tao Liu$^{1,2}$, Yu-Yu Zhang$^{1,*}$, Qing-Hu Chen$^{3}$, and Ke-Lin Wang$^{5}$}
\address{
$^{1}$ Department of Physics, Chongqing University, Chongqing 400044, P. R. China\\
$^{2}$Department of Physics, Southwest University of  Science and Technology, Mianyang 621010, P. R. China\\
$^{3}$Department of Physics, Zhejiang University, Hangzhou 310027,
P. R. China\\
$^{5}$Department of Modern Physics, University of  Science and
Technology of China,  Hefei 230026, P. R. China
 }

\thanks{* Corresponding author.}
\email{yuyuzh@cqu.edu.cn}

\date{\today}
\begin{abstract}
 Although the oscillator strength sum rule forbids the phase transition in ideal non-interacting two-level atoms systems, we present the possibility of the quantum  phase transition in the coupled two-level atoms in a cavity. The system undergoes the superradiant phase transition in the thermodynamics limit and this transition is account for the atom-atom attractive interaction, exhibiting a violation of the sum rule. The bosonic coherent state technique has been adopted to locate the quantum critical point accurately in the finite-size system. We predict the existence of the superadiant phase transition as the number of atoms increases, satisfying all the constraints imposed by the sum rule.
\end{abstract}



\maketitle
\section{introduction}

Successes in the control of two-level atoms coupled to cavities quantum electrodynamics (QED), especially the experimental realization of a circuit QED system operating in the ultrastrong-coupling limit~\cite{niem,wallraff,saidi,hartmann,simmonds,ciuti,devoret,forn}, are paving the way for the realization and the study of quantum phases for quantum applications.  As is known, the Dicke hamiltonian~\cite{dicke} as a model describing $N$ two-level atoms coupled to a single bosonic mode cavity exhibits a superrandiant phase transition~\cite{lieb,Emary,brandes,chen}. The order parameter of such a transition may be accessible by tuning the atom-cavity coupling across a critical point.
However, the classical supperradiant phase transition is forbidden by the inclusion of the squared electromagnetic vector potential $A^2$ in the Dicke Hamiltonian as a consequence of the Thomas-Reiche-Kuhn (TRK) sum rule for the oscillator strength~\cite{krza,bial,rza,currie}. Hence, a prediction of the phase transition is challenged by the corresponding sum rule.

For the case of real atoms in cavities, such as one-dimensional arrays of small Josephson junctions~\cite{barbara}, an ensemble of quantum dots~\cite{scheibner} and Bose-Einstein condensates~\cite{schbeble}, interacting between themselves and with the quantum field of cavities, the possibility of quantum phase transition, which occurs at zero temperature, is an attractive one. Recently, the quantum critical point is predicted in circuit QED system consisting of a collection of Josephson atoms capacitively coupled to a resonator due to a different topology with respect to real atoms, verifying the violation of the sum rule~\cite{nataf,nagy}. We construct additional example which show that quantum phase transition can occur in coupled two-level system and satisfy all the constraints imposed by the sum rule. The model, so-called a modified Dicke hamiltonian, have direct interatomic interactions in addition to the proper $A^2$ term.  An intriguing idea to take into account the interactions among atoms has recently attracted significant interests as many quantum phenomena closely related to the coupled-atoms emerge~\cite{gang,ball,chen1}. Since there are some works addressed the possibility of the classical phase transition in the Dicke model including $A^2$ terms and atom-atom interactions with different methods in infinite-size atoms system~\cite{Emeljanov,Keeling,sung}. It will be novel to address the existence of quantum phase transition both for finite and infinite two-level atoms by our extended bosonic coherent-state and Holstein-Primakoff approaches.

In this paper, the extended Dicke model including (a) atom-atom interactions, (b) interactions between atoms and a single-mode cavity, and (c)the $A^2$ term is considered. We present a numerical exact solution to the extended Dicke hamiltonian for finite atoms by the bosonic coherent state approach, which has been successfully employed to solve the related two-level systems~\cite{chen,chen1,zhang}. We obtain the critical point and thus can predict the possibility of the phase transition.
 We aim at the detailed predictions of the quantum phase transition by considering the interatomic interaction and $A^2$ term with the sum rule restriction in the thermodynamics limit and in the finite-size system.

 The paper is organized as follows: The hamiltonian for our model is introduced in Section II. In Section III, we presents the analytical quantum critical point in the thermodynamic limit and the TRK sum rule inequality. The possibility of quantum phase transition in finite-size system is discussed by a numerical exact solution in the section IV. A brief conclusion and discussion are given in the last section.

\section{the extended Dicke hamiltonian}

The Hamiltonian to describe the interaction between a collection of $N$
identical two-level atoms and the field of a cavity can be expressed as
\begin{equation}  \label{ham1}
H = H_{cav} + H_a + H_{cav-a} + H_{a-a} + H_{A^2}.
\end{equation}
Considering a single harmonic oscillator model of frequency $\omega$, the
field can be expressed by $H_{cav}=\omega a^{+}a$, where $a^{+}(a)$ is the
corresponding bosonic photon creation (annihilation) operator. $N$ identical
atoms are described as two-level system with the transition of frequency  $H_a=\frac \Delta 2\sum_{i=1}^N\sigma _z^i$ and $\sigma
_k^i(k=x,y,z)$ is the Pauli matrix for the $i$-th atom.

The interaction of the boson with the atoms with the coupling strength $\lambda $ is
\begin{equation}
H_{cav-a}=\frac \lambda {\sqrt{N}}(a^{+}+a)\sum_{i=1}^N\sigma _x^i,
\end{equation}
Note that we have include the counter-rotating terms. In the case of real atoms, a sufficient value of the coupling strength $\lambda$ is~\cite{krza}
\begin{equation}\label{lambda}
\lambda = \Delta d (\frac{2\pi\hbar}{\omega})^{1/2}(\frac{N}{V})^{1/2},
\end{equation}
where $d$ is a projection of the transition dipole moment on the polarization vector of the filed mode
and $V$ is the volume of the system with the atom density $\rho=N/V$.

Effects caused by interatomic interaction have been included by adding interaction term between the two-level atoms~\cite{gang,saidi,ball,chen1}
\begin{equation}
H_{a-a}=\frac \Omega {2N}\sum_{i\neq j}^N(\sigma _x^i\sigma _x^j+\sigma
_y^i\sigma _y^j),
\end{equation}
where $\Omega >0$ or $%
\Omega <0$ corresponds to the repulse or attractive interaction strength.

To account for the squared electromagnetic vector potential $A^2$ caused by
the longitudinal part of the bosonic filed, the quantized hamiltonian part
is
\begin{equation}
H_{A^2}=\kappa (a^{\dag}+a)^2,
\end{equation}
with $\kappa $ the oscillator strength. For $\lambda$ given in Eq.(~\ref{lambda}),
$\kappa$ in real atoms corresponds to
\begin{equation}\label{kapp}
\kappa=\frac{e^2}{2m}\frac{2\pi\hbar}{\omega}\rho.
\end{equation}

The Hamiltonian can be simplyfied in the Dicke representation
\begin{eqnarray}  \label{ham2}
H&=&\omega a^{+}a+\Delta J_z+\frac{2\lambda }{\sqrt{N}}(a^{+}+a)J_x
+\frac{2\Omega }N(J^2-J_z^2-\frac N2)+\kappa (a^{+}+a)^2,
\end{eqnarray}
with $J_k=\frac 12\sum_{i=1}^N\sigma _k^i$ . The state of the atomic part
lies in the subspace spanned by the Dicke states $\{|j,m\rangle
,m=-j,-j+1,...,j\}$ , which are known as the eigenstates of $J^2$ and $J_z$:
$J^2|j,m\rangle =j(j+1)|j,m\rangle $ and $J_z|j,m\rangle =m|j,m\rangle $.

Using an unitary transformation $e^{-iJ_y\pi /2}$, the extended Dicke model
is written as
\begin{eqnarray}  \label{ham3}
H&=&\omega a^{+}a-\Delta J_x+\frac{2\lambda }{\sqrt{N}}(a^{+}+a)J_z
+\frac{2\Omega }N(J^2-J_x^2-\frac N2)+\kappa (a^{+}+a)^2.
\end{eqnarray}
To diagonalize the $A^2$ term by means of the Bogoliubov transformation, we
introduce new bosonic operators
\begin{equation}
b^{+}=\mu a+\nu a^{+},b=\mu a^{+}+\nu a,
\end{equation}
with the coefficients satifying $\mu ^2-\nu ^2=1$. It can evaluated the
transformed hamiltonian by neglecting a constant as
\begin{eqnarray}  \label{ham4}
H=\omega _kb^{+}b-\Delta J_x+\frac{2\lambda _k}{\sqrt{N}}(b^{+}+b)J_z+\frac{
2\Omega }N(J^2-J_x^2),
\end{eqnarray}
where the effective parameters are $\omega _k=\omega \gamma _k$ and $\lambda
_k=\lambda /\sqrt{\gamma _k}$ with $\gamma _k=\sqrt{1+4\kappa /\omega }$.

\begin{figure}[tbp]
\includegraphics[width=8.8cm]{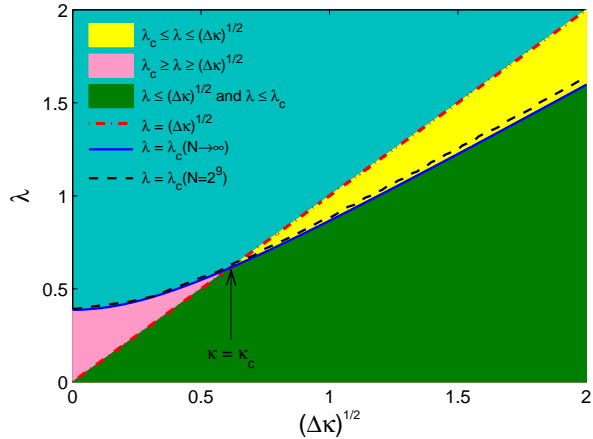}
\caption{\label{phase1}The phase diagram in the $(\Delta\kappa)^{1/2}-\lambda$ plane involving the atom-atom coupling and $A^2$ term. In constructing it we fix $\omega=1$ and $\Omega=-0.2$. The solid blue line, dashed red line are the critical coupling strength $\lambda=\lambda_c (N\rightarrow\infty)$ in the thermodynamics limit and $\lambda=(\Delta\kappa)^{1/2}$ by the TRK sum rule respectively, separating the super-radiant (upper left) and normal phases (lower right). The critical coupling strength $\lambda_c$ for $N=2^9$ two-level atoms is shown in dashed black line. In the yellow regions $\lambda_c<\lambda\leq (\Delta\kappa)^{1/2}$, there exists the superradiant phase transition.}
\end{figure}

\section{thermodynamics limit}
In order to show a quantum phase transition in the thermodynamic limit, we need to
find out the critical point and calculate the ground-state energy. The
thermodynamic limit is defined as $N\rightarrow \infty $ and $V\rightarrow
\infty $, while the atom density $\rho \propto N/V$ is kept constant.

With the Holstein-Primakoff mapping,  the angular momentum operators can be mapped into the single bosonic operators:
$J_{+}=c^{+}\sqrt{N-c^{+}c}$, $J_{-}=\sqrt{N-c^{+}c}c$ and $
J_z=c^{+}c-N/2$ with $[c,c^{+}]=1$. In order to describe the system above
the phase transition, we replace the bosonic mode operator $A^{+}$ and $B^{+}$ with
scaled auxiliary parameters $\alpha $ and $\beta $ as $A^{+}=b^{+}+\sqrt{N}
\alpha $ and $B^{+}=c^{+}-\sqrt{N}\beta $. After doing these replacements, the
approximated ground-state energy in Eq.(~\ref{energy}) can be expanded as:
\begin{eqnarray}\label{energy}
E_0/N =\omega _k\alpha ^2+\Delta (\beta ^2-\frac 12)-4\lambda _k\alpha
\beta \sqrt{1-\beta ^2}
-2\Omega (\beta ^2-\frac 12)^2+\frac \Omega 2.
\end{eqnarray}
The corresponding auxiliary parameters $\alpha $ and $\beta $ are derived
from minimizing the ground-state energy by $\partial E_0/\partial \alpha =0$
and $\partial E_0/\partial \beta =0$, which gives the critical point
\begin{equation}\label{critical}
\lambda _c=\sqrt{(\Delta +2\Omega )(\omega +4\kappa )}/2.
\end{equation}
There is a crucial
relation between the atom-atom coupling strength $\Omega$ and the oscillator strength $\kappa $
originating from the squared of the field $(a^{\dagger }+a)^2$, which
determine the existence of a superradiant quantum critical point.

The present extended Dicke Hamiltonian, involving the interactions among atoms and the $A^2$ terms, describes a system
undergoes a quantum phase transition from normal phase to supperradiant phase at the critical point $\lambda _c$. Above the quantum critical point, the ground state undergoes a dramatic change with a character of super-radiance, in which the atomic ensemble spontaneously emits with an intensity proportional to the squared of the number of atoms $N^2$ rather than $N$. It indicates that the super-radiant and normal phases are separated by the quantum critical coupling strength $\lambda_c$.

With the atom-cavity coupling strength $\lambda$ in Eq.(~\ref{lambda}) and the bosonic field coupling $\kappa$
in Eq.(~\ref{kapp}), it was addressed that the existence of a classical phase transition at finite temperature only if $d^2\Delta>e^2\hbar/2m$,
which contradicts to the TRK sum rule for the atoms, i.e., $d^2\Delta>e^2\hbar/2m$~\cite{krza}.
It can be evaluated that the sum rule requires
\begin{equation}\label{sum-rule}
\lambda <\lambda _c^{*}=\sqrt{\Delta \kappa }.
\end{equation}
And the restriction is strong enough to be important as we take into account the $A^2$ term.
We obtain the critical point $\lambda _c^{*}$ associated with the oscillator strength $\kappa$ due to the consideration of the sum rule in realistic system. Therefore, we can predict the existence of the QPT when the atom-cavity coupling strength satisfies $\lambda_c<\lambda<\lambda _c^{*}$.

\begin{figure}[tbp]
\includegraphics[width=8.5cm]{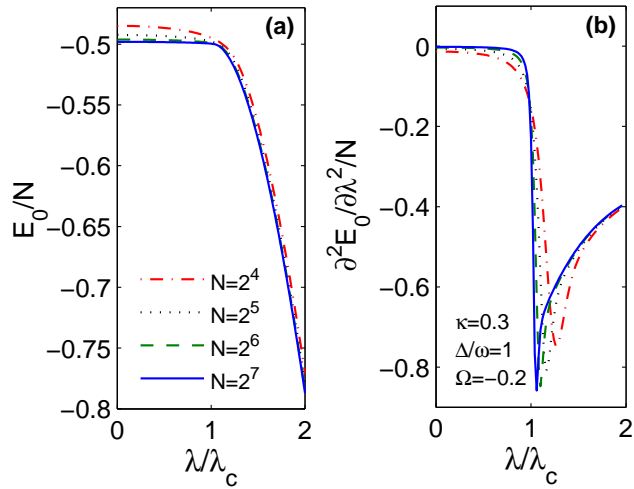}\label{critical point}
\caption{The scaled ground-state energy $E_0/N$ (a) and its second derivative $N^{-1}\partial^2E_0/\partial\lambda^2$ (b) as a function of coupling $\lambda/\lambda_c$ for $N=2^4,2^5,2^6,2^7$ atoms.}
\end{figure}
As a benchmark, we first address the generalized Dicke hamiltonian without interatomic interaction by setting $\Omega=0$. As is expected, the superradiant phase exists above the critical point $\lambda_c=\sqrt{\Delta(\omega+4\kappa)}/2$ obtained from eq. (~\ref{critical}). However, it is in contrast with the general sum rule inequality $\sqrt{\Delta(\omega+4\kappa)}/2<\sqrt{\Delta \kappa}$. In
such a no-go limit, the superradiant phase transition is forbidden.
Obviously, there does not exist critical regions where the atom-cavity coupling strength is required as $\lambda _c\leq\lambda\leq\sqrt{\Delta \kappa}$, resulting the impossibility of the quantum phase transition. In view of this, the inclusion of the $A^2$ term in the Dicke hamiltonian is verified to lead to the disappearance of quantum criticality as a consequence of the oscillator strength $\kappa$ sum rule.

In the case of interaction among atoms, $\Omega\neq0$, it is worthwhile to investigate the limitation for the supperadiant quantum phase transition. By looking into equations (~\ref{critical}) and (~\ref{sum-rule}), it is clear that it may be accessible to observe the quantum phase transition in particular coupling region
\begin{equation}
\sqrt{(\Delta +2\Omega )(\omega +4\kappa )}/2<\lambda<\sqrt{\Delta\kappa}.
\end{equation}
It is similar to the previous work by Bowden et al.~\cite{sung}, giving the conditions for the existence of the phase transition by a different method.
And the particular oscillator strength $\kappa$ are restricted as
\begin{equation}
\kappa>\kappa_c=\frac{\omega}{4}(\frac{\Delta}{2|\Omega|}-1).
\end{equation}
It is crucial for the existence of a superradiant quantum critical region: the relation of the oscillator strength $\kappa$, the atom-atom coupling strength $\Omega$ and the transition of frequency of the two-level atoms $\Delta$.
Consequencely, to predict the existence of the quantum phase transition, the interactions among atoms is restricted by the attractive interaction as
\begin{equation}
-\Delta/2<\Omega<0.
\end{equation}

We plot the phase diagram $(\Delta\kappa)^{1/2}-\lambda$ in the thermodynamics limit $N\rightarrow\infty$,
as shown in Fig.~\ref{phase1}. There are cross regions between the quantum critical coupling strength $\lambda=\lambda_c (N\rightarrow\infty)$ in solid blue line and $\lambda=(\Delta\kappa)^{1/2}$ by the TRK sum rule in dashed red line (see yellow region in Fig.~\ref{phase1}). It can predict the existence of the super-radiant phase transition in
the critical coupling region $\lambda_c<\lambda<\sqrt{\Delta\kappa}$ with the
atom-atom attractive coupling strength $ \Omega=-0.2$. It indicates that the no-go theorem is circumvented and the presence of $A^2$ term in the model hamiltonian does not lead to vanishing of the criticality due to the interatomic attractive coupling.

\begin{figure}[tbp]
\includegraphics[width=8.5cm]{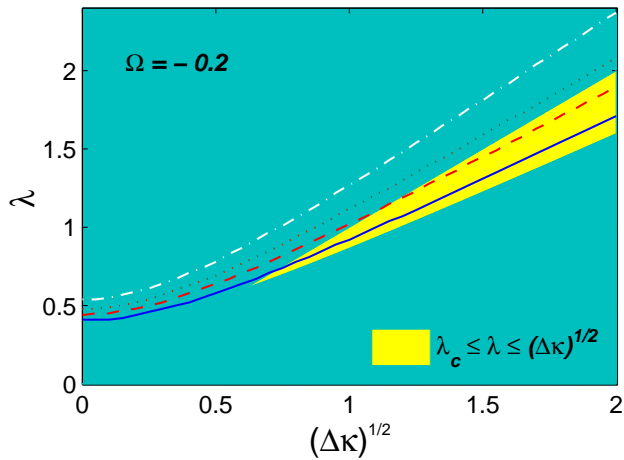}\label{phase2}
\caption{The phase diagram for finite-size atoms in the $(\Delta\kappa)^{1/2}-\lambda$ plane involving the atom-atom coupling and $A^2$ term. The dashed-dotted white line, dotted black line, dashed red line and solid blue line, are the phase boundaries for $\lambda=\lambda_c(N=2^4,2^5,2^6,2^7)$. As $ N$ increases to $2^6$, it enters into the yellow regions $\lambda_c<\lambda\leq (\Delta\kappa)^{1/2}$ and the super-radiant transition tends to occur. We choose $\omega=1$ and $\Omega=-0.2$.}
\end{figure}
\section{finite-size two-level atoms system}
In order to investigate the possibility of the quantum phase transition
for the finite-size atoms system, we first address the accurate solution to the extended Dicke model
by the bosonic coherent-state approach~\cite{chen,chen1,zhang}.
Since an exact solution for the extended Dicke Hamiltonian (~\ref{ham4}) in a finite system does not exist.
To perform an accurate numerical diagonalization, we
use the basis $\{|n\rangle _{A_m}\otimes |j,m\rangle \}$, where $|n\rangle
_{A_m}$ are position-displaced Fock states of the field and $|j,m\rangle $
are Dicke states. Note that $|n\rangle _{A_m}$ are the basis of the
replaced operator $A_m^{\dagger }=a^{\dagger }+g_m$ with $g_m=2\lambda
_km/(\omega _k\sqrt{N})$
\begin{equation}
|n\rangle _{A_m}=(A_m^{\dagger })^ne^{-g_ma^{\dagger }-g_m^2/2}|0\rangle ,
\end{equation}
by which the bosonic state of the Dicke state $|j,m\rangle $ can be
expressed as
\begin{equation}
|\varphi _m\rangle =\sum_n c_{n,m}|n\rangle _{A_m}.
\end{equation}
To carry out the diagonalization, we truncate the new bosonic Hilbert space
of the pseudospin accurately, which exhibits more efficient and better
convergence of the ground-state energy than the results obtained from the
diagonalization in Fock state, especially for large system sizes~\cite{chen,chen1}.
In this new basis, the $m$'th row of the Schr\"{o}dinger equation is expressed as:
\begin{eqnarray}
&\omega _k&(A_m^{\dagger }A_m-g_m^2)|\varphi _m\rangle |j,m\rangle +\frac{2N}
\Omega [j(j+1)
-j_{m-1}^{+}j_m^{-}-j_{m+1}^{-}j_m^{+}]|\varphi _m\rangle |j,m\rangle
-\Delta j_m^{-}|\varphi _m\rangle |j,m-1\rangle   \nonumber \\
&-&\Delta j_m^{+}|\varphi _m\rangle |j,m+1\rangle -\frac{2\Omega }
Nj_m^{-}j_{m-1}^{-}|\varphi _m\rangle |j,m-2\rangle
-\frac{2\Omega }Nj_m^{+}j_{m+1}^{+}|\varphi _m\rangle |j,m+2\rangle
=E|\varphi _m\rangle |j,m\rangle
\end{eqnarray}

The eigenvalues and eigenfunctions can be obtained accurately by exact diagonalization
with an accuracy less than $10^{-6}$, which facilitates the calculation for large system size~\cite{chen}.
Fig.~\ref{critical point} plots the scaled ground-state energy $E_0/N$ and the second derivative of the ground-state energy $N^{-1}\partial^2E_0/\partial\lambda^2$ for different atom-cavity coupling strength $\lambda/\lambda_c$ when atoms range from  $N=2^4$ to $N=2^7$.
The singularity at $\lambda_c$ clearly indicates the quantum critical point.

Fig.~\ref{phase1} shows that the critical coupling strength for $N=2^9$ atoms in dashed black line is much close to $\lambda_c$ in the thermodynamics limit, which indicates the numerical accurate results by the bosonic coherent approach. And the superradiant phase transition does exist in the yellow regions. Morever, Fig.~\ref{phase2} shows the critical coupling strength $\lambda_c$ for $N=2^4,2^5,2^6,2^7$ atoms
as a function of $(\Delta\kappa)^{1/2}$. It is obviously that there are no phase transition for $N=2^4,2^5$ due to the oscillator sum rule restriction. As $N$ increases to $2^6$, the critical coupling $\lambda_c$ enters into the yellow regions and the system undergoes the super-radiant phase transition, evidencing the counter-example of no-go theorem in finite-size system.

\section{conclusion}

In conclusion, detailed predictions of the possibility of the superradiant phase transition in the extended Dicke model are investigated by considering the atom-atom interaction and $A^2$ term with the TRK sum rule restriction. By the Holstein-Primakoff mapping in the thermodynamics limit, the analytical critical point $\lambda_c$ obtained contradicts to the sum rule restriction, which rules out the quantum phase transition in the absence of the interactions among atoms. However, attributing to the atom-atom attractive coupling $-\Delta/2<\Omega<0$, the no-go theorem is circumvented and there exist the quantum critical regions giving by $\lambda_c<\lambda<\sqrt{\Delta\kappa}$, which is associated with the oscillator strength $\kappa$ originating from the squared of the field $(a^{\dagger }+a)^2$. Morever, we solve the extended Dicke hamiltonian for finite atoms system by the  exact diagonalization with bosonic coherent state technique and the critical coupling strength is located accurately. For finite-size atoms system, it is possible to occur the superradiant phase transition as the number of atom $N$ increases.
With the consideration of interatomic interaction and the $A^2$ term, we show that such no-go theorem does not hold true attributing to the coupled two-level atoms and it is possible to occur superradiant quantum phase transition. The original properties may shed light on the realization of interesting quantum phases for quantum applications in cavity QED.

\acknowledgments This work is supported by National Natural Science Foundation of China under Grant Nos.11174254(QHC) and 11104363 (YYZ), National Basic Research Program of China under Grant No.2009CB929104 (QHC), and the Fundamental Research Funds for the Central Universities under Grant No. CDJZR11300006(YYZ). Addendum by authors in 2021 August£º Obviously, the  critical point $\lambda_c$  can be reduced by the attractive
interaction strength $\Omega$ and the negative $A^2$ term according to Eq. (\ref{critical}).


\begin{references}
\bibitem{niem} T. Niemczyk, F. Deppe, H. Huebl, E. P. Menzel, F. Hocke, M. J. Schwarz, J. J. Garcia-Ripoll, D. Zueco, T. H\"{u}mmer E. Solano, A. MARX, and R. Gross, Nature. Phys. \textbf{ 6}, 772(2010).
\bibitem{wallraff} A. Wallraff, D. I. Schuster, A. Blais, L. Frunzio, R. S. Huang, J. Majer, S. Kumar, S. M.Girvin, and R. J. Schoelkopf, Nature. Phys. \textbf{431}, 162(2004).
\bibitem{saidi} W. A. Al-Saidi, and D. Stroud, Phys. Rev. B \textbf{65}, 014512(2002);Phys. Rev. B\textbf{65}, 224512(2002).
\bibitem{hartmann}  M. J. Hartmann, et al., Nature. Phys. \textbf{2}, 849(2006);
A. D. Greentree, et al., ibid. 2,856(2006); D. G. Angelakis, M. F. Santos, V. Yannopapas, and
S. Bose, Phys. Rev. A \textbf{76},031805(2007).
\bibitem{simmonds}  R. W. Simmonds, et al., Phys. Rev. Lett.  \textbf{93}, 077003(2004).
\bibitem{ciuti}  C. Ciuti, and I. Carusotto, Phys. Rev. A  \textbf{74}, 033811(2006).
\bibitem{devoret}  M. Devoret, S. Girvin, and R. Schoelkopf, Ann. Phys. \textbf{16}, 767(2007).
\bibitem{forn}  P. Forn-D\'{i}az, et al., Phys.  Rew. Lett. \textbf{105}, 237001(2010).
\bibitem{dicke} R. H. Dicke, Phys. Rev. \textbf{ 93}, 99(1954).
\bibitem{lieb} K. Hepp and E. Lieb, Ann. Phys., \textbf{ 76}, 360(1973)
\bibitem{Emary} C. Emary and T. Brandes, Phys. Rev. E \textbf{67}, 066203(2003)
\bibitem{brandes} C. Emary and T. Brandes, Phys. Rev. Lett. \textbf{ 90}, 044101(2003).
\bibitem{chen}  Q. H. Chen, Y. Y. Zhang, T. Liu, and K. L. Wang,   Phys. Rev. A \textbf{78}, 051801(R)(2008).
\bibitem{krza}  K. Rza\'{z}ewski, K. W\'{o}dkiewicz, and W. \'{Z}acowicz  Phys. Rev. Lett \textbf{35}, 432(1975).
\bibitem{bial}  I. Bialynnicki-Birula, K. Rza\'{z}ewski, Phys. Rev. A \textbf{19}, 301(1979);
\bibitem{rza}  K. Rza\'{z}ewski, K. W\'{o}dkiewicz, Phys. Rev. Lett. \textbf{96},089301(2006).
\bibitem{currie}  E. Hadjimichael, W. Currie, and S. Fallieros,   Am. J. Phys. \textbf{65}, 335-341 (2010).

\bibitem{barbara}  P. Barbara, A. B. Cawthorne, S. V. Shitov, and C. J. Lobb, Phys. Rev. Lett. \textbf{82}, 1963(1999); B. Vasili\'{c}, P. Barbara, S. V. Shitov, and C. J. Lobb, Phys. Rev. B \textbf{65}, 180503(2002).
\bibitem{scheibner}  M. Scheibner, et al., Nature. Phys. \textbf{3}, 106(2007).
\bibitem{schbeble}  D. Schbeble, et al., Science. \textbf{300}, 475(2003).
\bibitem{nataf}  P. Nataf, and C. Ciuti,   Nature. Commun. \textbf{1}, 72 (2010).
\bibitem{nagy}  D. Nagy, G. K\'{o}nya, G. Szirmai, and P.Domokos, Phys. Rev. Lett. \textbf{104}, 130401 (2010).
\bibitem{gang}  G. Chen, X. G. Wang, J. Q. Liang,  and Z. D. Wang,   Phys. Rev. A  \textbf{78}, 023634(2008).
\bibitem{ball}  A. Ballesteros, O. Civitarese, F. J. Herranz, and M. Reboiro, Phys. Rev. B \textbf{68}, 214519(2003).
\bibitem{chen1}  Q. H. Chen, T. Liu, Y. Y. Zhang,  and K. L. Wang,   Phys. Rev. A  \textbf{82}, 053841(2010).
\bibitem{Emeljanov}  V. I. Emeljanov, and Y. L. Klimontovich, Phys. Lett. A 59, 366 (1976),
\bibitem{Keeling}  J. Keeling, J. Phys. Condensed matter : 19 295213 (2007),
\bibitem{sung}  C. M. Bowden,  and C. C. Sung,  J. Phys. A: Math. Gen, 11 151 (1978),

\bibitem{zhang}  Y. Y. Zhang, Q. H. Chen, and K. L. Wang,   Phys. Rev. B  \textbf{81}, 121105(R)(2010).


\end{references}
\end{document}